\documentclass[runningheads]{llncs}
\usepackage[T1]{fontenc}
\usepackage{newtxtext}
\usepackage[varvw]{newtxmath}
\usepackage{url}

\begin{document}
\title{Energy as a Service-Level Objective:\\Transaction-Level Energy Attribution}
\titlerunning{Energy as a Service-Level Objective}
\author{Julian Legler\orcidID{0000-0001-8715-0156}\\[0.4em]
{\normalsize supervised by Stefan Tai}}
\authorrunning{Julian Legler}
\institute{Technische Universit\"at Berlin, Berlin, Germany\\
Information Systems Engineering\\
\email{julian.legler@tu-berlin.de}}
\maketitle

\begin{abstract}
Service engineers assess every change against functional requirements and qualities such as reliability, performance and availability.
Energy is rarely part of this assessment.
When consumption rises after a change, the team cannot tell whether the increase is worth investigating, worth accepting, or neither.
A change to tracing, retries or a shared component can affect consumption far beyond the modified component.
This thesis asks how energy objectives can become actionable engineering instruments rather than reporting metrics.
Site Reliability Engineering (SRE) answered a similar question for reliability with service-level indicators, objectives and response policies.
We take SRE as the guiding research vision and study where the analogy holds for energy and where it breaks.
An \textit{energy objective} could, for example, bound the energy per \textit{successful transaction} under stated workload conditions.
The thesis connects three threads into one experimentally evaluated methodology.
Useful-work semantics define what a successful transaction is.
Energy attribution assigns shared and auxiliary consumption to transactions.
Service-level engineering decisions act on the resulting energy objectives.
Existing work covers attribution methods, measurement frameworks and sustainability objectives, mostly at component level and without a link to decisions.
Our prior work on service-level energy modeling showed that the energy of auxiliary services is often overlooked and hard to measure.
Planned experiments compare aggregate, request-normalized and useful-work-based indicators under controlled changes to observability configuration and injected failures.
Whether transaction-level attribution is needed for these decisions, or whether service-level attribution with outcome counts is sufficient, is an open question the experiments are designed to answer.
\end{abstract}

\section{Introduction}
Service engineers make decisions about code, configuration and architecture every day.
Regression tests assess these changes against selected functional requirements or system qualities, but a test suite detects only the properties it observes.
Consider a team increasing its tracing sampling rate to improve observability.
The change can generate additional computing, network traffic and storage overhead in the application and its observability backend.
In a deployment where the additional work leaves latency and availability within their objectives, those indicators alone would not indicate an energy regression.
Empirical studies of monitoring overhead show an energy and performance impact that is often overlooked~\cite{dingaEmpiricalEvaluationEnergy2023,leglerServiceLevelEnergyModeling2025}.

Detecting that energy consumption has increased is not enough.
What information would let the team judge whether better observability justifies it, identify who can respond, and evaluate the response?
Reducing observability may reduce energy consumption while adding blind spots.
An energy indicator therefore has to be read together with the service-level objectives, the workload and the trade-offs against other quality attributes of the whole system.

Reliability engineering faced similar issues in the past.
Site Reliability Engineering (SRE) addressed them by defining service-level objectives and linking them to operational decisions and responsibilities~\cite{beyer2018sreworkbook}.
We therefore take SRE as the guiding research vision.
This thesis investigates how service-level indicators (SLIs), service-level objectives (SLOs) and response policies connect energy measurements to decisions.
We focus on how useful-work accounting with shared energy attribution compares with simpler request-normalized metrics such as energy per request.
The literature already offers methods that produce energy signals and methods that consume signals for service-level decisions, but the two are not connected.
The contribution of this thesis is a methodology that connects useful-work semantics, energy attribution and service-level engineering decisions, and that is evaluated experimentally.
The expected results show when energy objectives become actionable engineering instruments rather than reporting metrics.
In particular, they show when transaction-level energy attribution is needed for a decision and when simpler indicators are sufficient.

\section{Problem and Research Questions}
\label{sec:problem}
An energy-efficient refactoring may not reduce total energy consumption if the request rate rises in the same period.
Lower energy per request can in turn hide additional attempts needed to complete the same user task.
Metrics like total energy, energy per service, energy per server, energy per attempt, energy per request, energy per hour and energy per accepted outcome all answer different questions.
Without information about the intended outcome and the overall workload, each of them can be misleading.

We use \emph{transaction} as a working term for an application-level unit of activity or work started by a user or another system.
It may span multiple services.
We cut the accounting boundary at the organization and report work done by services of other organizations as unattributed, because we see at most the spans they export and have no energy data for them.
Where to draw the accounting boundary within an organization, in particular for idle energy and supporting systems, is one of the key open questions for our model.

A \emph{successful transaction} is our working term for the result of useful work.
It is the completion of the intended activity in a way that meets the application's functional and quality requirements, for example correct data returned within a time limit.
Who decides what counts as a successful transaction is an open question, and the answer may differ by use case.
A retry is an attempt toward a successful transaction, not a new transaction.

An \emph{aborted or cancelled transaction} is a transaction that was started but did not complete successfully, either because the user cancelled it or because the system failed.
Failed attempts with multiple retries therefore add energy but are not counted as successful transactions.

\emph{Energy per successful transaction} is the currently planned indicator for measuring the energy efficiency of service-oriented systems.
An energy SLO then bounds this indicator, for example as a maximum energy per successful transaction under stated workload conditions.
Section~\ref{sec:method} discusses the candidate forms of such an objective.
Whether a successful transaction can be derived from traces, or whether a product owner has to declare the criteria per activity, is an open question.
We expect that the criteria have to be declared per workflow and that no general rule exists, but we have not tested this.
A further hypothesis is that useful work can only be judged across multiple services, because a single service rarely knows whether the user's intent was fulfilled.
This is why we speak of successful transactions rather than successful requests.
A further open question is how to handle idle energy and the energy of supporting systems such as observability backends, and whether this energy should be allocated to transactions or reported separately.

The model therefore needs an energy allocation policy that states how shared consumption is divided among transactions or services and how idle energy and supporting systems are treated.

Our overarching question is \emph{how energy objectives can become actionable engineering instruments rather than reporting metrics}.
Its technical core is under which conditions successful transactions, shared-energy allocation and estimation uncertainty change energy-regression judgments and intervention choices in service-oriented systems.
We break it down into three research questions.
\begin{description}
\item[RQ1:] How can successful transactions and accounting boundaries be specified for service activities with retries, failures and supporting work?
\item[RQ2:] How sensitive are attributed values and regression judgments to allocation policy, workload mix and estimation uncertainty?
\item[RQ3:] How can transaction-aware energy indicators and service-level objectives guide engineering decisions, and under which conditions are simpler indicators sufficient?
\end{description}

\section{Related Work}

Existing work already defines useful work in the context of energy metrics.
The Green Software Measurement Model (GSMM) by Guldner et al. treats useful work as application-specific \cite{guldnerGSMM2024}.
It only allows a comparison when both systems do comparable work and produce equivalent outcomes.
The Software Carbon Intensity (SCI) specification, standardized as ISO/IEC 21031, defines operational and embodied carbon per functional unit, with energy as one of its inputs \cite{gsf2024sci}.
All components of a system should use the same functional unit.
SCI leaves the functional unit deliberately open so that it fits different use cases.
Neither GSMM nor SCI gives a concrete method to decide which outcomes count as useful or how to allocate shared costs in service-oriented systems.

Other works attribute energy to individual transactions.
Our own GOXN work models the CPU, memory, network and storage energy of each service, including auxiliary services \cite{leglerServiceLevelEnergyModeling2025}.
Brunnert and Gutzy extend the OpenTelemetry Java agent (OTJAE) to estimate the energy of a transaction from its resource activity and to compute SCI per transaction \cite{brunnertGutzyGreenMetrics2024}.
Spanergy by Batista et al. splits the energy of a service equally among the spans that run at the same time \cite{batista_spanergy_2026}.
This keeps the total of all active spans intact, but energy consumed outside these spans is not attributed to anyone.

Sharma and Fuerst split the energy of the full system into serverless function invocations and include shared hardware and software costs \cite{sharmaFaasMeter2024}.
Schneider and Mattia allocate idle energy by reservation and dynamic energy by usage, and then pass the cost of shared services on to their consumers along the service chain \cite{schneiderCloudCarbon2024}.
Anand et al. show trade-offs between components and propose energy budgets per request \cite{anandOddOneOut2023}.
This supports end-to-end accounting, but it does not provide a general outcome definition or allocation rule.
Each of these approaches makes different accounting decisions.
As a result, the footprint of a transaction depends on the measurement boundary and on the allocation policy for shared consumption.
We have not decided whether to build on one of these methods or to implement our own attribution on top of GOXN.

The SRE Workbook connects SLOs to error-budget policies that name responses and responsible people \cite{beyer2018sreworkbook}.
This process is our research vision.

GreenSLAs by Botero et al. express the energy consumption of a service as an SLO and connect the contract to flexibility requests, rewards and penalties \cite{botero2013greenslas}.
CASCA by Herrera et al. connects telemetry and declared SLO ranges to reconfiguration at runtime \cite{herrera2026casca}.
In their media-streaming evaluation, encoding resources are adjusted based on quality and sustainability signals.
Vitali et al. break an annual carbon budget down into hourly energy budgets \cite{vitaliAdaptiveGreen2026}.
Within these budgets, they choose component versions, scaling and optional features to balance user experience and revenue.
Their simulation uses modeled energy rather than measurements, and they do not show how such policies affect the decisions of real service teams.

\section{Research Approach and Methodology}
\label{sec:method}
SRE uses error budgets to balance the pace of change with system reliability~\cite{sloss2016sreintroduction}.
What we want to transfer is a process with explicit objectives, measurement windows, response policies and clear responsibilities.
Whether this works for energy and where the analogy breaks is a hypothesis of this research.
The comparison with a reliability SLO does not hold in every respect.
A reliability SLI is often a ratio of good events to valid events.
It is bounded between 0 and 100\% and describes a symptom that users notice.
An energy measurement has neither property by itself.
Energy has no natural ceiling and is invisible to users.
It is also unclear whether an energy budget should reset, accumulate, or be expressed as a threshold instead.
Who should set the energy objective is an open question.

We currently see three candidate forms for an energy SLO.
The first is an intensity threshold, for example a maximum energy per successful transaction evaluated over a window.
This form keeps the window and the response policy of SRE.
The second is a constraint on energy efficiency under stated workload conditions.
The third is a violation budget, an allowed number of windows per period in which the threshold is exceeded.
This form is structurally the closest to an error budget, because it turns energy back into a count of good and bad windows.
What makes a window good or bad still has to be decided.
Which form is suitable, and whether an energy budget is a useful concept at all, is part of RQ3.

Clear definitions of successful transactions and energy objectives should also clarify which role is responsible for which response.

\paragraph{RQ1.}
We plan to define successful transaction criteria for selected workflows of a microservice system and to record attempts, failures and completion states in controlled experiments.
Our GOXN tooling currently runs the OpenTelemetry astronomy-shop demo on Kubernetes and measures energy with Kepler.
We plan to add further Systems Under Experimentation (SUE) and workloads to check how far the results generalize.
We use distributed tracing to join spans into transactions and service-level measurements to add shared and auxiliary energy.
Whether we continue the tracing sampling rate experiments started in GOXN is open.

\paragraph{RQ2.}
We plan to compare indicators such as total energy, energy per attempt and energy per successful transaction under different transaction-attribution methods.
The experiments vary one factor at a time.
One such factor is the retry behavior under injected failures.
Workload, request mix and acceptance criteria stay fixed within a trial.
Later experiments can use other SUEs, other workloads and more complex transactions.

We then implement transaction-level attribution in this setup and compare it against service-level attribution.
We check that the attributed values add up to the aggregate measurements and that interventions produce measurable changes.
The goal is to separate the attributable from the non-attributable part of the consumption and to show that the attributed values stay consistent with the system-level measurements.

\paragraph{RQ3.}
We again vary one factor at a time and keep all other conditions constant.
The goal is to see how each factor changes the decision made under the declared quality constraints, and whether transaction-level attribution leads to different decisions than service-level attribution.
One option is to set an energy budget and observe how the attribution method changes the decisions taken to stay within it.

Detecting a controlled regression under measurement noise will probably be the main challenge of this evaluation.
This will then hopefully demonstrate the practical utility of the attribution methods in guiding energy-efficient decisions under realistic conditions.

\section{Preliminary and Ongoing Work}
Our published GOXN work provides a service-level energy model that covers CPU, memory, network, storage and auxiliary services, together with tooling for experiments~\cite{leglerServiceLevelEnergyModeling2025}.
In the evaluated configurations, leaving out network and storage underestimated the energy of auxiliary services by up to 63\%.
Much of the energy of an observability change sits outside the service call graph, in storage and tracing backends.
The GOXN experiments also showed considerable measurement noise and variability in this environment, which complicates attribution to individual services and actions.
Careful experimental design and repeated measurements still produced usable results.

READ, our ongoing work on attributing GPU energy to spans in LLM inference, confirmed two observations that carry over to this thesis.
Attribution policies differ in how they treat idle energy, and fairness of an allocation and its physical accuracy are separate questions.
Shared inference itself is out of scope for this thesis.

\section{Research Plan and Discussion}
This paper is the first written formulation of this research agenda with SRE as the guiding research vision.
The idea grew out of the observations and challenges of the GOXN paper published at ICSOC 2025~\cite{leglerServiceLevelEnergyModeling2025} and a personal interest in SRE.
In the next six months we want to settle the outcome and energy accounting semantics for the astronomy-shop workflows, extend GOXN with transaction-level attribution and establish repeatable baselines (RQ1).
The allocation policy and indicator comparison (RQ2) can start as soon as the baselines are stable and will overlap with the first phase.
The objective and response policy evaluation (RQ3) depends on the outcome of RQ2 and is the least settled part.
Additional SUEs beyond the astronomy-shop demo are planned to check generalizability.
A practitioner study via an ongoing research project (Software Campus) is an optional extension and not required for the thesis.

The main risks we currently see are limited access to measurements across service boundaries, attribution models that are complex and hard to validate, and limited generality across workflows and platforms because the definition of a successful transaction varies.

\clearpage
\bibliographystyle{splncs04}
\bibliography{references}
\end{document}